\documentclass[showpacs,showkeys,preprintnumbers,amsmath,amssymb]{revtex4}

\usepackage{graphicx}
\usepackage{dcolumn}
\usepackage{bm}

\begin{document}
\renewcommand{\thesection}{\arabic{section}}
\renewcommand{\thetable}{\arabic{table}}
\bibliographystyle{apsrev}

\title{\center{On the geometric potential derived from
 \\  Hermitian momenta on a curved surface}}

\author{M. Encinosa }

\affiliation{ Florida A\&M University \vskip1pt Department of
Physics \vskip1pt Tallahassee FL 32307}



\begin{abstract}
A geometric potential $V_C$ depending on the mean and Gaussian
curvatures of a surface $\Sigma$ arises when confining a particle
initially in a three-dimensional space $\Omega$ onto  $\Sigma$
when the particle Hamiltonian $H_\Omega$  is taken proportional to
the Laplacian $L$ on $\Omega$. In this work rather than assume
$H_\Omega \propto L$, momenta  $P_\eta$ Hermitian over $\Omega$
are constructed and used to derive an alternate Hamiltonian
$H_\eta$. The procedure leading to $V_C$, when performed with
$H_\eta$,  is shown  to yield $V_C = 0$. To obtain a measure of
the difference between the two approaches, numerical results are
presented for a toroidal model.

\end{abstract}

\maketitle

\section{Introduction}

The study of quantum mechanics on curved surfaces has been a
subject of
  theoretical effort for decades \cite{jenskoppe,dacosta1,dacosta2,exnerseba,
 matsutani,burgsjens,duclosexner,bindscatt,goldjaffe,
  ouyang,popov,midgwang,clarbrac,schujaff,ee1,
ee2,lin,chapblic,qu,nils}. The approach as pioneered in
\cite{burgsjens, jenskoppe, dacosta1,dacosta2} first defines a
surface $\Sigma(u,v)$ and employs a coordinate $q$ to describe
excursions at any point normal to $\Sigma$. The Hamiltonian
appropriate to the three dimensional coordinate system $\Omega$
characterized by a metric $g_{ij}$ is then defined by $H_\Omega
\equiv -{1\over2} \nabla^2$ with\
$$
\nabla^2= g^{-{1 \over 2}}{\partial \over \partial q^i} \bigg [
g^{1 \over 2}\ g^{ij}{\partial \over \partial q^j} \bigg ].
\eqno(1)
$$
\vskip 3pt As will be briefly reviewed below, Eq.(1), along with
imposing conservation of the norm in the limiting case where the
particle is constrained to motion on $\Sigma$, leads to a
geometric potential $V_C$. This paper with is concerned with how
$V_C$ is modified when instead of defining $H_\Omega$ through
Eq.(1), another choice of Hamiltonian, one for which surface and
normal momenta are required to be Hermitian over $\Omega$ is
adopted.

The remainder of this paper is organized as follows: in section 2
the geometric potential is derived from Eq.(1) for a cylindrically
symmetric geometry $\Omega_c$. This material has appeared other
places, but is included here to keep this work as self contained
as possible and to establish some notation. A symmetric geometry
was chosen for ease of presentation; the salient points for more
general cases remain intact. In section 3 the classical
Hamiltonian for a particle in $\Omega_c$ is written, Hermitian
momenta on $\Omega_c$ are found, and standard quantum mechanical
rules  employed to write the first quantized Hamiltonian. Finally,
using the limiting procedure detailed in section 2, the geometric
potential is re-derived. Section 4 presents numerical calculations
for a toroidal system, and section 5 is reserved for conclusions.

\section{The geometric potential; a brief review.}
The derivation of $V_C$ will be performed for a cylindrically
symmetric geometry $\Omega_c$ with $H_\Omega \propto L$. The
extension to the general case is straightforward and the central
results below will be not be affected by the restriction to
azimuthal symmetry.

Let ${\mathbf e}_{\rho},{\mathbf e}_{\phi}$ be cylindrical
coordinate system unit vectors defined through
$$
\rm {\mathbf e}_{\rho}=cos\phi \ {\mathbf i}+ sin \phi \ {\mathbf
j} \eqno(2)
$$
$$
\rm {\mathbf e}_{\phi}=-sin \phi \ {\mathbf i}+ cos \phi \
{\mathbf j} \eqno(3)
$$
 and let a cylindrically symmetric surface $\Sigma_c$ be described
  by
$$
{\mathbf r}(\rho,\phi)=\rho   {\mathbf e}_{\rho} + S(\rho)
{\mathbf k}. \eqno(4)
$$
 $S(\rho)$ gives the shape of the surface. Any point near
 $\Sigma_c$ may be reached by travelling normal to $\Sigma_c$ up (or down) a distance $q$
 via
$$
{\mathbf x}(\rho,\phi,q)={\mathbf r}(\rho,\phi)+q  {\mathbf e}_{n}
\eqno(5)
$$
\noindent with ${\mathbf e}_{n}$ everywhere normal to the surface.
The metric near the surface derived from Eq. (5) is \cite{ee1}
(subscripts here indicate differentiation)
$$
dx^2=  Z^2 \bigg[ 1-{{qS_{\rho\rho}} \over {Z^3}} \bigg ]^2
d\rho^2 + \rho^2 \bigg[ 1-{{qS_{\rho}} \over {\rho Z}} \bigg ]^2
d\phi^2 +\ dq^2 \eqno(6)
$$
$$
\equiv Z^2 [ 1+q k_1  ]^2  d\rho^2 + \rho^2 [ 1+qk_2]^2 d\phi^2 +\
dq^2 \eqno(7)
$$
with $Z=\sqrt {1+S_\rho^2}$. The Laplacian can be found
straightforwardly from Eq. (7) but is not essential to the main
arguments that follow.

Now  consider a situation where a large confining potential
$V_n(q)$ everywhere normal to $\Sigma_c$ acts to restrict the
particle to the surface. The derivation of the geometric potential
is independent of its specific form \cite{scripta}, but however
chosen it causes the particle to sit on $\Sigma_c$. In this limit
the wave function decouples into surface and normal parts
\cite{jenskoppe,dacosta1,dacosta2,burgsjens}
$$
\Psi(\rho,\phi,q) \rightarrow \chi_s(\rho,\phi)\chi_n(q). \eqno(8)
$$
Conservation of the norm in the decoupled limit is imposed by
insisting upon the condition
$$
|\Psi|^2 F d\Sigma dq=|\chi_s|^2|\chi_n|^2d\Sigma dq \eqno(9)
$$
 with $F=1+2qh+q^2k$ and $d\Sigma$ the surface measure.  $h$ and $k$  are the mean and Gaussian curvatures given by
$$
h={1\over 2}(k_1+k_2), \eqno(10)
$$
$$
k=k_1k_2. \eqno(11)
$$
Inserting the expression given by Eq.(8)
$$\Psi = {\chi_s\chi_n \over \sqrt F}
\eqno(12)
$$
 into the time independent Schrodinger equation $-{1\over 2}\nabla^2\Psi = E\Psi$, and taking
the $q\rightarrow 0$ limit after performing all $q$
differentiations, yields the pair of equations $(\hbar = m = 1)$
$$
-{1\over 2}\bigg[
 {1\over Z^2}{\partial^2 \over \partial \rho^2}
 +{1\over Z^2\rho}{\partial \over \partial \rho}
 -{Z_{\rho}\over Z^3}{\partial  \over \partial \rho}
 +{1\over  \rho^2}{\partial^2 \over \partial \phi^2}
 +(h^2-k)\bigg]\chi_s  =E_s
\chi_s \eqno(13)
$$
$$
-{1 \over 2}{\partial^2  \chi_n \over \partial q^2} + V_n(q)
\chi_n =E_n \chi_n. \eqno(14)
$$
The curvature potential $V_C$ is
$$
V_C = -{1 \over 2}[h^2-k]; \eqno(15)
$$
seeing specifically how $V_C$ is generated will prove important in
the following section. It arises from the $q \rightarrow 0$ limit
of the
$$
\bigg ( {\partial^2 \over \partial q^2} +  2h {\partial \over
\partial q}\bigg )
\chi_s(\rho,\phi)\chi_n(q) F^{-{1/2}} \eqno(16)
$$
\vskip 3pt \noindent term in the full Laplacian . The $Z$
dependence appearing in the kinetic energy
 operator given by Eq. (13) occurs independently of the $q$ degree of freedom.
The effect on the spectra of some simple systems due to metric
factors in the operator was shown to be small in \cite{ee1,ee2}
for some parameterizations of $S(\rho)$, but it is also possible
for the effects to be large \cite{encunpub}.

\section{The Hamiltonian with Hermitian momenta}

In this section a Hamiltonian $H_\eta$ quadratic in momenta
$P_\eta$ Hermitian over $\Omega_c$ is developed.  The procedure
followed comprises standard quantum mechanical rules, to wit:

\noindent a. the classical Lagrangian is written \vskip 3pt
\noindent b. the canonical momenta are determined\vskip 3pt
\noindent c. the classical momenta are replaced with their first
quantized form  $\hat{P}_\eta$ Hermitian on $\Omega_c$. \vskip 6pt

\noindent The  Lagrangian for a particle in a geometry described
by Eq. (6)  is
$$
L = {1\over 2} \big(Z^2[1+k_1q]^2 \dot{\rho}^2+ \rho^2[1+k_2q]^2
\dot{\phi}^2+\dot{q}^2 \big). \eqno(17)
$$
With $P_\eta = {\partial L / \partial \dot{q}_\eta}$, the classical
Hamiltonian for this geometry is
$$
H = {1\over 2} \bigg( {P_\rho^2 \over Z^2[1+k_1q]^2} + {P_\phi^2
\over \rho^2[1+k_2q]^2} + P_q^2 \bigg). \eqno(18)
$$
To proceed with first quantization via $P_\eta \rightarrow
\hat{P}_\eta$, it is necessary to produce momentum operators
Hermitian on a geometry described by $g_{ij}$. The required
relation is
$$
\hat{P_\eta}={1\over i} \bigg( {\partial \over \partial q_\eta} +
{1 \over 2}{\partial \over \partial q_\eta}{\rm ln} \sqrt g
\bigg). \eqno(19)
$$
Anticipating the $q \rightarrow 0$ limit, it is useful to note the
differentiations done by the surface momenta will pass over any
appearances of the $q$ variable, making it possible to set $q = 0$
in both of them without affecting the final result. The momenta in
this limit are
$$
\hat{P}_\rho={1\over i} \bigg( {\partial \over
\partial \rho} + {1 \over 2}{\partial Z \over \partial \rho} + {1 \over 2\rho}
\bigg)\eqno(20)
$$
$$
\hat{P}_\phi={1\over i} {\partial \over
\partial \phi} \eqno(21)
$$
$$
\hat{P}_q={1\over i} \bigg( {\partial \over
\partial q} + {1 \over F}[h+qk]
\bigg). \eqno(22)
$$
$H_\eta$ is straightforward to obtain from the three momenta
 listed above, but it is Eqs. (16) and
(22) that will be focused upon to generate the central result of
this paper. Write
$$
-\hat{P}_q^2= {\partial^2 \over \partial q^2}+ {2\over F}
(h+qk){\partial \over\partial q}+ {1 \over F} \bigg( {\partial h
\over \partial q} +k \bigg)- {1\over F^2} {\partial F \over
\partial q} (h+qk) + {1\over F^2} (h+qk)^2;
 \eqno(23)
$$
 upon taking the $q \rightarrow 0$ limit the result is
$$
-\hat{P}_q^2= {\partial^2 \over \partial q^2}+ 2h {\partial
\over\partial q}+ k - h^2.
 \eqno(24)
$$
The key point is simple; Eq.(24) includes the operator that
appears in the Laplacian formulation of the problem as shown in
Eq.(16). It eventually yields $\partial^2 \over \partial q^2$ plus
$h^2 - k$, from which it is trivially seen that the geometric
potential cancels in the Hermitian formulation.

\section{A numerical example}
In this section the spectra and wave functions of the Schrodinger
equation for a particle on a toroidal surface $T^2$ of major
radius $R$ and minor radius $a$ will be compared using the two
formalisms detailed above. There are good reasons to choose such a
system, not the least being  calculating system observables is
tractable. Additionally, toroidal nanostructures of varied types
have been fabricated so there is some relevance to real devices, a
point which will be elaborated upon in the conclusions section
below.

The surface of a doughnut shaped torus has may be parameterized
 by \cite{fpl}
$$
 {\mathbf x}(\theta,\phi,q)=(R + a \ {\rm cos}  \theta ){\mathbf e}_{\rho} +a\  {\rm sin}
\theta{\mathbf {k}} + q {\mathbf e}_{n}. \eqno(25)
$$
 Applying $d$ to Eq.(25)
gives
$$
d{\mathbf x}= (a+q)d\theta  {\mathbf e}_{\theta}+[R + (a+q) \ {\rm
cos} \theta]d\phi {\mathbf e}_{\phi}+dq {\mathbf e}_n \eqno(26)
$$
with ${\mathbf e}_{\theta} =-{\rm sin} \theta  {\mathbf
e}_{\rho}+{\rm cos}\theta {\mathbf k}$ and ${\mathbf e}_{n}\equiv
{\bf e}_{\phi} \  {\rm x} \  {\mathbf e}_{\theta}$.

Let $\alpha = a/R$, $\beta = 2Ea^2$; applying the formalism
described in section 2 and making the standard separation ansatz
for the azimuthal part of the wave function $\Psi(\theta,\phi) =
\psi(\theta)exp \ [i\nu\phi]$ yields the Schrodinger equation with
$V_C$ present \cite{encmott},
$$
  {\partial^2 \psi \over \partial \theta^2} -
 {\alpha \ {\rm sin}\ \theta \over [1 + \alpha \   \cos\theta]}{\partial \psi \over \partial \theta}
-{(\nu^2 \alpha^2- {1\over4}) \over [1 + \alpha \
\cos\theta]^2}\psi +\beta\psi = 0. \eqno(27)
$$
The method of section 3 gives the Schrodinger equation
$$
  {\partial^2 \psi \over \partial \theta^2} -
 {\alpha \ {\rm sin}\ \theta \over [1 + \alpha \   \cos\theta]}{\partial \psi \over \partial \theta}
-{[\nu^2 \alpha^2+ {1 \over 4}(\alpha^2 - 1)] \over [1 + \alpha \
\cos\theta]^2}\psi +(\beta-{1\over 4}\psi) = 0. \eqno(28)
$$
\vskip 6pt \noindent Even before consideration of the solutions of
Eq. (27) and (28), there is an immediate distinction between the
two. The cancellation of the azimuthal kinetic energy term at
discrete values of $\alpha$  occurs  in the former  at values of
$$
\alpha = {1 \over 2\nu} \eqno(29)
$$
and in the latter at
$$
\alpha = \sqrt{{1 \over 1+4\nu^2}}. \eqno(30)
$$
\vskip 6pt \noindent
 The ramifications of Eq.(29) and by implication Eq.(30) have been
discussed elsewhere. Here it need be noted that while there
appears a  term in the Hermitian formalism that is identical to
$V_C$ derived from the Laplacian method, it is an artifact of the
high degree of symmetry of the torus and may take a very different
form in the general case.

Numerical solutions of Eqs. (27) and (28) were determined by a
basis set expansion in Gram-Schmidt functions on $T^2$
\cite{encgs}. Eigenvalues and wave functions are given in tables
1-3 for several low-lying states with $\alpha=1/3, 1/2$ and $2/3$.
The results indicate clearly that the two Hamiltonians can differ,
particularly at the \lq\lq magic" $\alpha$ given by Eq. (29) or
Eq. (30), and when the  torus fattens to values of $\alpha
\thicksim 2/3$.

\section{Conclusions}
This work demonstrated that a Hamiltonian quadratic in momenta
Hermitian on a curved surface and consistent with standard quantum
mechanical rules leads to a vanishing geometric potential $V_C$.
It is interesting that Golovnev has recently employed an alternate
(Dirac quantization) procedure for which $V_C =0$ on a sphere
\cite{golovnev}; however, it is not clear to the author how the DQ
method could be easily implemented on $T^2$. Numerical results
were presented to show the differences in spectra and wave
functions between the two Hamiltonians discussed here can be
substantial. It is possible to speculate as to whether advances in
nanostructure fabrication may perhaps allow for experimental
determination of which  (if either) Hamiltonian studied here is
the more correct description of curved surface nanophysics. Given
that
 toroidal structures have been fabricated and are \lq\lq clean" with
 respect to curvature effects (in that there is enough symmetry to make the
  problem interesting but not so much to make it trivial),
it is conceivable that existing devices may  be employed to settle
the very fundamental issue of what is the appropriate Hamiltonian
for a particle constrained to a surface.  \vskip 6pt

\section{Acknowledgements}
The author would like to thank Mr. Lonnie Mott for useful
discussions.

\newpage
\bibliography{hermbib}

\begin{thebibliography}{26}
\expandafter\ifx\csname natexlab\endcsname\relax\def\natexlab#1{#1}\fi
\expandafter\ifx\csname bibnamefont\endcsname\relax
  \def\bibnamefont#1{#1}\fi
\expandafter\ifx\csname bibfnamefont\endcsname\relax
  \def\bibfnamefont#1{#1}\fi
\expandafter\ifx\csname citenamefont\endcsname\relax
  \def\citenamefont#1{#1}\fi
\expandafter\ifx\csname url\endcsname\relax
  \def\url#1{\texttt{#1}}\fi
\expandafter\ifx\csname urlprefix\endcsname\relax\def\urlprefix{URL }\fi
\providecommand{\bibinfo}[2]{#2}
\providecommand{\eprint}[2][]{\url{#2}}

\bibitem[{\citenamefont{Jensen and Koppe}(1971)}]{jenskoppe}
\bibinfo{author}{\bibfnamefont{H.}~\bibnamefont{Jensen}} \bibnamefont{and}
  \bibinfo{author}{\bibfnamefont{H.}~\bibnamefont{Koppe}},
  \bibinfo{journal}{Ann. of Phys.} \textbf{\bibinfo{volume}{63}},
  \bibinfo{pages}{586} (\bibinfo{year}{1971}).

\bibitem[{\citenamefont{da~Costa}(1981)}]{dacosta1}
\bibinfo{author}{\bibfnamefont{R.~C.~T.} \bibnamefont{da~Costa}},
  \bibinfo{journal}{Phys. Rev. A} \textbf{\bibinfo{volume}{23}},
  \bibinfo{pages}{1982} (\bibinfo{year}{1981}).

\bibitem[{\citenamefont{da~Costa}(1982)}]{dacosta2}
\bibinfo{author}{\bibfnamefont{R.~C.~T.} \bibnamefont{da~Costa}},
  \bibinfo{journal}{Phys. Rev. A} \textbf{\bibinfo{volume}{25}},
  \bibinfo{pages}{2893} (\bibinfo{year}{1982}).

\bibitem[{\citenamefont{Burgess and Jensen}(1993)}]{burgsjens}
\bibinfo{author}{\bibfnamefont{M.}~\bibnamefont{Burgess}} \bibnamefont{and}
  \bibinfo{author}{\bibfnamefont{B.}~\bibnamefont{Jensen}},
  \bibinfo{journal}{Phys. Rev. A} \textbf{\bibinfo{volume}{48}},
  \bibinfo{pages}{1861} (\bibinfo{year}{1993}).

\bibitem[{\citenamefont{Exner and Seba}(1989)}]{exnerseba}
\bibinfo{author}{\bibfnamefont{P.}~\bibnamefont{Exner}} \bibnamefont{and}
  \bibinfo{author}{\bibfnamefont{P.}~\bibnamefont{Seba}}, \bibinfo{journal}{J.
  Math. Phys.} \textbf{\bibinfo{volume}{30}}, \bibinfo{pages}{2574}
  (\bibinfo{year}{1989}).

\bibitem[{\citenamefont{Matusani}(1991)}]{matsutani}
\bibinfo{author}{\bibfnamefont{S.}~\bibnamefont{Matusani}},
  \bibinfo{journal}{J. Phys. Soc. Jap.} \textbf{\bibinfo{volume}{61}},
  \bibinfo{pages}{55} (\bibinfo{year}{1991}).

\bibitem[{\citenamefont{Duclos and Exner}(1995)}]{duclosexner}
\bibinfo{author}{\bibfnamefont{P.}~\bibnamefont{Duclos}} \bibnamefont{and}
  \bibinfo{author}{\bibfnamefont{P.}~\bibnamefont{Exner}},
  \bibinfo{journal}{Rev. Math. Phys.} \textbf{\bibinfo{volume}{7}},
  \bibinfo{pages}{73} (\bibinfo{year}{1995}).

\bibitem[{\citenamefont{Londergan et~al.}(1999)\citenamefont{Londergan, Carini,
  and Murdock}}]{bindscatt}
\bibinfo{author}{\bibfnamefont{J.}~\bibnamefont{Londergan}},
  \bibinfo{author}{\bibfnamefont{J.}~\bibnamefont{Carini}}, \bibnamefont{and}
  \bibinfo{author}{\bibfnamefont{D.}~\bibnamefont{Murdock}},
  \emph{\bibinfo{title}{Binding and scattering in two dimensional systems;
  applications to quantum wires, waveguides, and photonic crystals}}
  (\bibinfo{publisher}{Springer-Verlag}, \bibinfo{address}{Berlin},
  \bibinfo{year}{1999}).

\bibitem[{\citenamefont{Goldstone and Jaffe}(1991)}]{goldjaffe}
\bibinfo{author}{\bibfnamefont{J.}~\bibnamefont{Goldstone}} \bibnamefont{and}
  \bibinfo{author}{\bibfnamefont{R.~L.} \bibnamefont{Jaffe}},
  \bibinfo{journal}{Phys. Rev. B} \textbf{\bibinfo{volume}{45}},
  \bibinfo{pages}{14100} (\bibinfo{year}{1991}).

\bibitem[{\citenamefont{Ouyang et~al.}(1998)\citenamefont{Ouyang, Mohta, and
  Jaffe}}]{ouyang}
\bibinfo{author}{\bibfnamefont{P.}~\bibnamefont{Ouyang}},
  \bibinfo{author}{\bibfnamefont{V.}~\bibnamefont{Mohta}}, \bibnamefont{and}
  \bibinfo{author}{\bibfnamefont{R.~L.} \bibnamefont{Jaffe}},
  \bibinfo{journal}{Ann. of Phys.} \textbf{\bibinfo{volume}{275}},
  \bibinfo{pages}{297} (\bibinfo{year}{1998}).

\bibitem[{\citenamefont{Popov}(2000)}]{popov}
\bibinfo{author}{\bibfnamefont{I.}~\bibnamefont{Popov}},
  \bibinfo{journal}{Phys. Lett. A} \textbf{\bibinfo{volume}{269}},
  \bibinfo{pages}{148} (\bibinfo{year}{2000}).

\bibitem[{\citenamefont{Midgley and Wang}(2000)}]{midgwang}
\bibinfo{author}{\bibfnamefont{S.}~\bibnamefont{Midgley}} \bibnamefont{and}
  \bibinfo{author}{\bibfnamefont{J.}~\bibnamefont{Wang}},
  \bibinfo{journal}{Aus. J. Phys.} \textbf{\bibinfo{volume}{53}},
  \bibinfo{pages}{77} (\bibinfo{year}{2000}).

\bibitem[{\citenamefont{Clark and Bracken}(1996)}]{clarbrac}
\bibinfo{author}{\bibfnamefont{I.~J.} \bibnamefont{Clark}} \bibnamefont{and}
  \bibinfo{author}{\bibfnamefont{A.~J.} \bibnamefont{Bracken}},
  \bibinfo{journal}{J. Phys. A} \textbf{\bibinfo{volume}{29}},
  \bibinfo{pages}{4527} (\bibinfo{year}{1996}).

\bibitem[{\citenamefont{Schuster and Jaffe}(2003)}]{schujaff}
\bibinfo{author}{\bibfnamefont{P.~C.} \bibnamefont{Schuster}} \bibnamefont{and}
  \bibinfo{author}{\bibfnamefont{R.~L.} \bibnamefont{Jaffe}},
  \bibinfo{journal}{Ann. Phys.} \textbf{\bibinfo{volume}{307}},
  \bibinfo{pages}{132} (\bibinfo{year}{2003}).

\bibitem[{\citenamefont{Encinosa and Etemadi}(1998{\natexlab{a}})}]{ee1}
\bibinfo{author}{\bibfnamefont{M.}~\bibnamefont{Encinosa}} \bibnamefont{and}
  \bibinfo{author}{\bibfnamefont{B.}~\bibnamefont{Etemadi}},
  \bibinfo{journal}{Phys. Rev. A} \textbf{\bibinfo{volume}{58}},
  \bibinfo{pages}{77} (\bibinfo{year}{1998}{\natexlab{a}}).

\bibitem[{\citenamefont{Encinosa and Etemadi}(1998{\natexlab{b}})}]{ee2}
\bibinfo{author}{\bibfnamefont{M.}~\bibnamefont{Encinosa}} \bibnamefont{and}
  \bibinfo{author}{\bibfnamefont{B.}~\bibnamefont{Etemadi}},
  \bibinfo{journal}{Physica B} \textbf{\bibinfo{volume}{266}},
  \bibinfo{pages}{361} (\bibinfo{year}{1998}{\natexlab{b}}).

\bibitem[{\citenamefont{Lin and Jaffe}(1996)}]{lin}
\bibinfo{author}{\bibfnamefont{K.}~\bibnamefont{Lin}} \bibnamefont{and}
  \bibinfo{author}{\bibfnamefont{R.~L.} \bibnamefont{Jaffe}},
  \bibinfo{journal}{Phys. Rev. B} \textbf{\bibinfo{volume}{54}},
  \bibinfo{pages}{5757} (\bibinfo{year}{1996}).

\bibitem[{\citenamefont{Chaplik and Blick}(2004)}]{chapblic}
\bibinfo{author}{\bibfnamefont{A.}~\bibnamefont{Chaplik}} \bibnamefont{and}
  \bibinfo{author}{\bibfnamefont{R.~H.} \bibnamefont{Blick}},
  \bibinfo{journal}{New J. Phys.} \textbf{\bibinfo{volume}{6}},
  \bibinfo{pages}{33} (\bibinfo{year}{2004}).

\bibitem[{\citenamefont{Qu and Geller}(2004)}]{qu}
\bibinfo{author}{\bibfnamefont{S.}~\bibnamefont{Qu}} \bibnamefont{and}
  \bibinfo{author}{\bibfnamefont{M.}~\bibnamefont{Geller}},
  \bibinfo{journal}{Phys. Rev. B} \textbf{\bibinfo{volume}{70}},
  \bibinfo{pages}{085414} (\bibinfo{year}{2004}).

\bibitem[{\citenamefont{Nilsson}()}]{nils}
\bibinfo{author}{\bibfnamefont{B.}~\bibnamefont{Nilsson}},
  \bibinfo{note}{\texttt{cond-mat/0103029}}.

\bibitem[{\citenamefont{Encinosa et~al.}(2005)\citenamefont{Encinosa, L.Mott,
  and Etemadi}}]{scripta}
\bibinfo{author}{\bibfnamefont{M.}~\bibnamefont{Encinosa}},
  \bibinfo{author}{\bibnamefont{L.Mott}}, \bibnamefont{and}
  \bibinfo{author}{\bibfnamefont{B.}~\bibnamefont{Etemadi}},
  \bibinfo{journal}{Phys. Scr.} \textbf{\bibinfo{volume}{72}},
  \bibinfo{pages}{13} (\bibinfo{year}{2005}).

\bibitem[{\citenamefont{Encinosa}({\natexlab{a}})}]{encunpub}
\bibinfo{author}{\bibfnamefont{M.}~\bibnamefont{Encinosa}},
  \eprint{unpublished}.

\bibitem[{\citenamefont{Encinosa and Etemadi}(2003)}]{fpl}
\bibinfo{author}{\bibfnamefont{M.}~\bibnamefont{Encinosa}} \bibnamefont{and}
  \bibinfo{author}{\bibfnamefont{B.}~\bibnamefont{Etemadi}},
  \bibinfo{journal}{Found. Phys. Lett.} \textbf{\bibinfo{volume}{16}},
  \bibinfo{pages}{403} (\bibinfo{year}{2003}).

\bibitem[{\citenamefont{Encinosa and L.Mott}(2003)}]{encmott}
\bibinfo{author}{\bibfnamefont{M.}~\bibnamefont{Encinosa}} \bibnamefont{and}
  \bibinfo{author}{\bibnamefont{L.Mott}}, \bibinfo{journal}{Phys. Rev. A}
  \textbf{\bibinfo{volume}{68}}, \bibinfo{pages}{014102}
  (\bibinfo{year}{2003}).

\bibitem[{\citenamefont{Encinosa}({\natexlab{b}})}]{encgs}
\bibinfo{author}{\bibfnamefont{M.}~\bibnamefont{Encinosa}},
  \bibinfo{note}{\texttt{physics/0501161}, submitted to App. Math. Comp.}

\bibitem[{\citenamefont{Golovnev}()}]{golovnev}
\bibinfo{author}{\bibfnamefont{A.}~\bibnamefont{Golovnev}},
  \bibinfo{note}{\texttt{quant-ph/0508044}}.

\end{thebibliography}

\newpage
\begin{table}
\caption{Eigenvalues and wave functions corresponding to the
solutions of Eqs. (27) and (28) for $\alpha ={1 \over 3}$ (the
$(2\pi)^{-{1\over2}}$ normalization from from the $\phi$
dependence is omitted.) Terms not shown are at least an order of
magnitude smaller than those listed.}
\begin{center}
\begin{tabular}{|l|l|l|}
\hline
\  $\quad \ \beta$ & \ \ \ $\alpha = {1 \over 3}$  \ \  $V_C \neq 0$ \\
\hline
\ -.2834 & \ $\Psi_{0}=.4082- 0.0776\rm cos \theta$\\
\ -.1528 & $\ \Psi_{1}=[.4049- 0.0421\rm cos \theta]e^{\pm i\phi}$  \\
\  +.1968 & $\ \Psi_{2}=[.4060- 0.0525\rm cos \theta]e^{\pm 2i\phi}$  \\
\hline
\  $\quad \ \beta$ & \ \ \  $\alpha = {1 \over 3}$  \ \  $V_C = 0$ \\
\hline
\ \ \ \ 0 &     $\ \Psi_{0}=-.4075- 0.0686\rm cos \theta$\\
\ +.1267 & $\ \Psi_{1}=[.4038- 0.0335\rm cos \theta]e^{\pm i\phi}$  \\
\ +.4735 & $\ \Psi_{2}=[.3869- 0.0593\rm cos \theta]e^{\pm 2i\phi}$  \\
 \hline
\end{tabular}
\end{center}
\end{table}

\begin{table}
\caption{As per table 1 with  $\alpha ={1 \over 2}$. The middle
state is qualitatively very different due to $\alpha ={1 \over 2}$
defining a \lq \lq magic radius" when $V_C = 0$. }
\begin{center}
\begin{tabular}{|l|l|l|}
\hline
\  $\quad \ \beta$ & \ \ \ $\alpha = {1 \over 2}$  \ \  $V_C \neq 0$ \\
\hline
\ -.3511 & \ $\Psi_{0}=.4230- 0.1470\rm cos \theta$\\
\ \ \ \ 0 & $\ \Psi_{1}= e^{\pm i\phi}$  \\
\  +.6288 & $\ \Psi_{2}=[-.3293- 0.1935\rm cos \theta]e^{\pm 2i\phi}$  \\
\hline
\  $\quad \ \beta$ & \ \ \  $\alpha = {1 \over 2}$  \ \  $V_C = 0$ \\
\hline
\ \ \ \ 0 &     $\ \Psi_{0}=.4191- 0.1072\rm cos \theta$\\
\ +.3192 & $\ \Psi_{1}=[.3912+0.0293\rm cos \theta]e^{\pm i\phi}$  \\
\ +.9212 & $\ \Psi_{2}=[-.3289- 0.2008\rm cos \theta]e^{\pm 2i\phi}$  \\
 \hline
\end{tabular}
\end{center}
\end{table}

\begin{table}
\caption{As per table 1 with  $\alpha ={2 \over 3}$. The second
excited state for the case when $V_C = 0$ becomes a negative
parity function in $\theta$.}
\begin{center}
\begin{tabular}{|l|l|l|}
\hline
\  $\quad \ \beta$ & \ \ \ $\alpha = {2 \over 3}$  \ \  $V_C \neq 0$ \\
\hline
\ -.5947 & \ $\Psi_{0}=0.4597- 0.3406\rm cos \theta-0.1000\rm cos 2 \theta$\\
\ +.2024 & $\ \Psi_{1}= [0.3570+0.1079\rm cos \theta]e^{\pm i\phi}$  \\
\  +.4377 & $\ \Psi_{2}=0.5930\rm cos \theta-0.1363cos 2\theta$  \\
\hline
\  $\quad \ \beta$ & \ \ \  $\alpha = {2 \over 3}$  \ \  $V_C = 0$ \\
\hline
\ 3.4$(10^{-4})$ & \ $\Psi_{0}=-0.4369+ 0.1516\rm cos \theta$\\
\ +.5397 & $\ \Psi_{1}= [0.3405+0.1446\rm cos \theta]e^{\pm i\phi}$  \\
\  +1.0000 & $\ \Psi_{2}=-0.5888\rm sin \theta-0.0992 sin 2\theta$  \\
 \hline
\end{tabular}
\end{center}
\end{table}

\end{document}